\documentclass{ws-procs961x669}            

\def \bea {\begin{eqnarray}}
\def \eea {\end{eqnarray}}
\def \nn   {\nonumber}

\begin{document}
\title{Minimal length discretization and properties of modified metric tensor and geodesics}
\author{Abdel Nasser Tawfik$^*$}
\address{Egyptian Center for Theoretical Physics (ECTP), Cairo, Egypt \\
World Laboratory for Cosmology And Particle Physics (WLCAPP), Cairo, Egypt
$^*$E-mail: atawfik@cern.ch}

\author{Fady T. Farouk, F. Salah Tarabia, Muhammad Maher}
\address{Helwan University, Faculty of Science, Physics Department, 11792 Ain Helwan, Cairo, Egypt}

\begin{abstract}
We argue that the minimal length discretization generalizing the Heisenberg uncertainty principle, in which the gravitational impacts on the non--commutation relations are thoughtfully taken into account, radically modifies the spacetime geometry. The resulting metric tensor and geodesic equation combine the general relativity terms with additional terms depending on higher--order derivatives. Suggesting solutions for the modified geodesics, for instance, isn't a trivial task. We discuss on the properties of the resulting metric tensor, line element, and geodesic equation.
\end{abstract}


\bodymatter

\section{Introduction}
\label{intro}

Suggesting a consistent theory for quantum gravity is an ultimate goal of recent research aiming at unifying quantum mechanics (QM) with the theory of general relativity (GR). The quantum nature of gravity is still an open question in physics. It turns to form an enigma occupying the minds of many physicists for many decades, for instance, so--far there is no scientific explanation for the origin of the gravitational fields and how particles behave inside them! This makes the gravity foundamentally different from the other fources of nature. There were various attempts to reconcile some principles of GR in a quantum framework \cite{Rovelli:1989za,Donoghue:1994dn}. The quantum gravity - on one hand - is conjectured to add new elements to GR and QM. On the other hand, the Einstein field equations combine classical geometry with quantum stress--energy--momentum tensor! For QM, we recall the possible modification in the Heisenberg uncertainty principle due to gravitational effects\footnote{They weren't taken into considration in the original Heisenberg uncertainty principle.}, known as generalized uncertainty princile (GUP) \cite{Tawfik:2014zca,Tawfik:2015rva}. The present work introduces some insights on the possible impacts of GUP on the metric tensor and the geodesic equation and their properties. 

Couple decades ago, the line element modification and the gravity quantization have been suggested in litrature \cite{Feoli:1999cn,Capozziello:2000gb,Bozza:2001qm}. A review of these models could be found in refs. \cite{Hess:book,Hess:2020ssc,Tawfik:2020zvf}. An extended class of the metric tensors which are functions of a field of an internal vectors $y^{\alpha}(x)$, $g_{\mu\nu}(x^{\alpha}\cdot y^{\alpha}(x))$, was also discussed in litrature \cite{Oliveira:1985cj}. Accordingly, the origin of the gravitational field of spin-1 massless particle could be determined \cite{Oliveira:1985cj}. Regarding the equivalence principle of gravitationan and enirtial mass in extended gravity and non--geodesic motion, a machian request was scratched \cite{Licata:2016qmz}, where a direct coupling between the Ricci curvature scalar and the matter Lagrangian could be fixed in  cosmological observations \cite{Licata:2014xgx}.

The present script aims at tackling the long--standing essential puzzle that the Einstein field equations relate {\it non--quantized} semi--Riemannian geometry expressed in Ricci and Einstein tensors, which are basically depending on the metric tensor $g_{\mu\nu}(x^{\alpha})$, with the {\it full--quantized} stress--energy-momentum tensor $T_{\mu\nu}$ \cite{Stephani:624239}. The approach applied here is the finite minimal length uncertainty obtained from GUP. GUP was inspired by black hole physics, doubly special relativity, and string theory \cite{Tawfik:2014zca,Tawfik:2015rva}. It is worthy highligting that GUP is  comparable with the Planck length, where quantum fluctuations in {\it quasi--quantized} manifold likely emerge \cite{Karolyhazy1966,Tawfik:2016uhs,Tawfik:2019dda}. 

The modifications in the metric tensor and geodesic equation due to GUP shall be introduced. Accordingly, we believe that the present study offers a theoretical framework for the observations that the universe acceleratingly expands. Alongside forces deriving this type of expansion, such as dark energy \cite{Tawfik:2019dda} and cosmological constant \cite{Peebles:2002gy}, we suggest that it might be also originated to interdependence of GR on QM, especially at the Planck scale.

The present paper is organized as follows. Section \ref{sec:gup} reviews the generalized uncertainly principle and minimal length uncertainty. The discretized metric tensor and its properties shall be discussed in section \ref{sec:st}. The discretized geodesic equation and its properties shall be introduced in section \ref{sec:ge}. Section \ref{sec:cncls} is devoted to the final conclusions.

\section{Generalized uncertainly principle and minimal length uncertainty}
\label{sec:gup}

The Heisenberg uncertainty principle (HUP) is a recipe on how the uncertainties are constrained in the quantum non--commutation relations of length and momentum operators, for instance. This was originally suggested in absence of any gravitational field. Analyzing HUP in gravitational fields comes up with modifications, know as GUP. Such a phenomenological aspect can be originated in black hole physics, doubly special relativity, and string theory \cite{Tawfik:2014zca,Tawfik:2015rva}. Phenomenology because so--far there is no theoretical explanation of the gravitational fields origin making gravity substantially different from the other three forces of nature. A finite minimal length was suggested as \cite{Kempf:1994su},
\bea
\Delta x\, \Delta p\geq \frac{\hbar}{2} \left[1+ \beta (\Delta p)^2 \right], \label{GUP}
\eea 
where $\Delta x$ and $\Delta p$ are length and momentum uncertainties, respectively. $\beta = \beta_0 (\ell_p/\hbar)^2 = \beta_0/ (m_p c)^2$, is the GUP parameter with $\ell_p=\sqrt{\hbar G/c^3}=1.977 \times 10^{-16}~$GeV$^{-1}$ the Planck length and  $m_p=\sqrt{\hbar c/G}= 1.22 \times 10^{19}~$GeV$/c^2$ the Planck mass. An upper bound on  $\beta_0$ is based on astronomical observations, such as, $\beta_0 \lesssim 5.5 \times 10^{60}$ based on recent gravitational waves observations. From Eq. (\ref{GUP}), then the minimum length uncertainty reads 
\bea
\Delta x_{\mbox{min}} \approx \hbar \sqrt{3\beta_0} &=& \ell_p \sqrt{\beta_0}.
\eea 
The experimental estimations \cite{Gao:2016fmk} of $\beta_0$ range from $\sim 10^{21}$ to $\sim 10^{39}$. Another GUP approach motivated by quantum deformation of the Pioncarre algebra \cite{Maggiore:1993rv,Maggiore:1993zu} suggests that the minimal length associated with is given as
\bea
\Delta x_{\mbox{min}} \simeq  \frac{\hbar G}{\gamma_0} &=& \frac{\ell_p^2}{\gamma_0},
\eea 
where $\gamma_0$ is photon's wavelength sent from infinity to be absorbed in a black hole. Its experimental estimations \cite{Gao:2016fmk} range from $\sim 10^{26}$ to $\sim 10^{34}$. 

The UV/IR correspondence can be applied on various phenomena in which short and long distance physics play a characterized role. For example, gravity \cite{Jackson:2005ue} and the {\it ''deformed''} commutation relations have features of UV/IR correspondence \cite{Jackson2005GravityFA}. Therefore, we assume that the minimal length discretization could be analyzed from the UV/IR correspondence \cite{Tawfik:2020zvf}. 
Indeed, $\Delta x$ rapidly increases (IR) as the $\Delta p$ increases beyond the Planck scale (UV) \cite{Maldacena:1997re,Gubser:1998bc,Witten:1998qj}.  

Analogous to expression (\ref{GUP}), the canonical non--commutation relation of length and momentum quantum operators is given as 
\bea
\left[\hat{x}_i\,, \hat{p}_j\right] \geq \delta_{ij} i\, \hbar \left(1 + \beta p^2 \right), \label{GUP2}
\eea
where $p^2=g_{ij} p^{0i}\, p^{0j}$ and $g_{ij}$ is Minkowski spacetime metric tensor, for instance $(-,+,+,+)$. Both length and momentum operators, respectively, read
\bea 
\hat{x}_i &=& \hat{x}_{0i} (1+\beta p^2),  \label{GUP3a}\\
\hat{p}_j &=& \hat{p}_{0j}, \label{GUP3b}
\eea
in which $\hat{x}_{0i}$ and $\hat{p}_{0j}$ can be derived from the corresponding non--communitation relation,
\bea
[\hat{x}_{0i}, \hat{p}_{0j}]&=&\delta_{ij} i\, \hbar.
\eea 

\section{Discretized metric tensor and its properties}
\label{sec:st}

With the quantum non--commutation operations discussed in section \ref{sec:gup} the four--dimensional Minkowskian manifold of the line element can be given as
\bea
ds^2 &=& g_{\mu\nu} dx^\mu dx^\nu, 
\eea
where $\mu$, $\nu$, and $\lambda=0, 1, 2, 3$. This could be extended to eight--dimensional spacetime tangent bundle with coordinates $x^A=(x^\mu(\zeta^a), \beta \dot{x}^\mu(\zeta^a))$, where $\dot{x}^\mu = dx^\mu/d\zeta^\mu$ \cite{Brandt:2000gka}. Obviously, the line element reads 
\bea
d \tilde{s}^{2} = g_{AB} \, dx^A \, dx^B, \label{eq:dS}
\eea
where $x^A= x^A(\zeta^\mu)$, $g_{AB} = g_{\mu_\nu}\otimes g_{\mu\nu}$, and the indices $A$ and $B=0, 1, \cdots, 7$. With some assumptions and approximations, this manifold could be reformulated as an effective four--dimensional spacetime manifold,
 \bea
 \tilde{g}_{\mu\nu} &=& g_{AB} \frac{\partial x^A}{\partial \zeta^\mu} \frac{\partial x^B}{\partial \zeta^\nu} \simeq g_{ab} \Big[\frac{\partial x^a}{\partial \zeta^\mu}  \frac{\partial x^b}{\partial \zeta^\nu} + \beta \frac{\partial \dot{x}^a}{\partial \zeta^\mu}  \frac{\partial \dot{x}^b}{\partial \zeta^\nu} \Big]  \simeq  \left(1+ \beta \ddot{x}^\lambda \ddot{x}_\lambda \right) g_{\mu\nu}.
 \eea
$\ddot{x}^\mu = \partial \dot{x}^\mu/\partial \zeta^\mu$ is four--dimensional acceleration.  The indices $a$, and $b=0, 1, \cdots, 7$.

It is now proper to highlight a number of the discretized metric tensor properties. First, we distinguish between flat and curved spacetime:
\begin{itemize}
\item For flat spacetime, $g_{\mu \nu} = \eta_{\mu\nu}$ and the modified metric tensor reads
\bea
\tilde{g}_{\mu\nu} &\simeq&   \eta_{\mu \nu} + \beta \ddot{x}^\lambda \ddot{x}_\lambda  \eta_{\mu \nu} \simeq \eta_{\mu \nu} + h_{\mu\nu},
\eea
where $h_{\mu\nu} = \beta \ddot{x}^\lambda \ddot{x}_\lambda \eta_{\mu \nu}$. The quantum fluctuations from which the minimal length uncertainty likely emerges are compassed in $h_{\mu\nu}$. In other words, both GR metric tensor and this modified one are only distinguishable, at finite $h_{\mu\nu}$.
\item For curved spacetime, the modified metric tensor reads
\bea
\tilde{g}_{\mu\nu} &\simeq &  g_{\mu\nu} + \beta \ddot{x}^\lambda \ddot{x}_\lambda  g_{\mu\nu}. \label{eq:dmunucurvedtilda}
\eea
Again, $\beta \ddot{x}^\lambda \ddot{x}_\lambda g_{\mu\nu}$ represents the minimal length contributions.
\end{itemize}

The $\tilde{g}_{\mu\nu}$, Eq. (\ref{eq:dmunucurvedtilda}), seems to play the same role as that of $g_{\mu\nu}$, namely it turns a covariant tensor into a contravariant tensor and vice versa. $\tilde{g}_{\mu\nu}$ is apparently symmetric in its indices $\mu$ and $\nu$ as its components are likely equal upon exchanging the indices. The $\tilde{g}_{\mu\nu}$ symmetric properties are invariant upon basis transformation only if its indices are of the same type; variant or covariant. For spacetime represented by a four--dimensional differentiable manifold $\mathcal{M}$, the metric tensor, $\tilde{g}_{\mu\nu}$, Eq. (\ref{eq:dmunucurvedtilda}), is covariant, second--degree, symmetric on $\mathcal{M}$. Such   manifold is Lorentzian. In local coordinates $x^{\mu}$, the metric can be expressed in $dx^{\mu}$ (one--form gradients of the scalar coordinate fields $x^{\mu}$), and the coefficients $\tilde{g}_{\mu\nu}$ ($16$ real--valued functions). If the local coordinates are specified, the metric can be written as a $4 \times 4$ symmetric matrix with the coefficients $\tilde{g}_{\mu\nu}$. If the matrix is non--singular having non--vanishing determinant, $\tilde{g}_{\mu\nu}$ is non--degenerate. The Lorentzian signature of $\tilde{g}_{\mu\nu}$ obviously implies that the matrix has one negative and three positive eigenvalues. Under a change of the local coordinates $x^{\mu}\to x^{\bar{\mu}}$, we obtain
\bea
\tilde{g}_{\bar{\mu}\bar{\nu}} = \frac{\partial x^{\alpha}}{\partial x^{\bar{\mu}}} \, \frac{\partial x^{\beta}}{\partial x^{\bar{\nu}}}\; \tilde{g}_{\alpha \beta} = \Theta^{\alpha}_{\bar{\mu}} \, \Theta^{\beta}_{\bar{\nu}}  \; \tilde{g}_{\alpha \beta} 
\eea

If $dx^{\mu}$ are regarded as the components of an infinitesimal coordinate displacement four--vector, the metric determines the invariant square of an infinitesimal line element. Also, it is apparent that the principle of the general covariance is satisfied even in  absence of gravitational effects on the modified Minkowski metric. Then, the modified four--dimensional line element can be expressed as
\bea
d \tilde{s}^{2} &=& g_{\mu\nu} \left( dx^\mu dx^\nu+ \beta^2  d\dot{x}^\mu \; d\dot{x}^\nu\right) = ds^2 + \beta^2 \ddot{x}^\lambda \ddot{x}_\lambda \, ds^2,
\eea
and $d \tilde{s}^{2}$ manifests the causal structure of the spacetime. Only timelike four--dimensional line element, $d \tilde{s}^{2}<0$, can physically covered by a massive object. If $d \tilde{s}^{2}=0$ the line element is lightlike (covered by massless photon). On the other hand, for $d \tilde{s}^{2}>0$, the four--dimensional line element is an incremental proper length, i.e., spacelike.

\section{Discretized geodesic equation and its properties}
\label{sec:ge}

The properties of the manifold including geodesics in GR [$\ddot{x}^{\mu}+\Gamma^{\mu}_{\alpha\beta} \dot{x}^{\mu}\dot{x}^{\nu}=0$ with $\dot{x}^{\mu}=\partial x^{\mu}/\partial s$, Christopher symbols $\Gamma^{\mu}_{\alpha\beta}=(g^{\mu\gamma}/2)(\partial g_{\gamma\alpha}/\partial x^{\beta}) + \partial g_{\gamma\beta}/\partial x^{\alpha} - \partial g_{\alpha\beta}/\partial x^{\gamma})$, and $s$ is a scalar parameter of motion such as a proper time] can be generalized. For example, the notion of a {\it ''straight line''} is to be generalized to {\it ''curved spacetime''}. The consequences of the length discretization or minimal length uncertainty based on GUP, generalize the world line of a free particle. GR assumes that gravity couldn't be regarded as a force but it should be emerged from {\it ''classical''} curved spacetime geometry and {\it ''quantized''} stress--energy--momentum tensor is causing the {\it ''classical''} spacetime curvature. The approach utilized here is discretized minimal length, which modifies both metric tensor and line element.

As discussed in ref. \cite{Tawfik:2020zvf}, using the variational principle and extremization of the path $s_{AB}$, the geodesic equation could be derived as follows.
\begin{itemize}
\item In flat spacetime (Minkowski space)
\bea
\beta {\cal L} \frac{d^2 \dot{x}^{\mu}}{d s^2} - \frac{d x^{\mu}}{d s} + c &=& 0.
\eea
Even if $\beta$, the GUP parameter vanishes, the geodesic equation reduces to that of a straight line, at constant velocity, i.e., both $\beta$ terms disappear.
\item and in curved spacetime
\bea
\frac{d^2 x^{\mu}}{d s^2} + \Gamma^{\eta}_{\mu \nu} \frac{d x^{\mu}}{d s} \frac{d x^{\nu}}{d s} &=&  \beta \Bigg\{  \frac{d}{d s} \Big[ {\cal L} \frac{d^2 \dot{x}^{\mu}}{d s^2} \Big] +\frac{1}{2}  g^{\mu \nu} \Bigg( 
  g_{\mu\nu, \alpha} \frac{d \dot{x}^{\mu}}{d s} \frac{d \dot{x}^{\nu}}{d s} -    {\cal L} g_{\mu \alpha, \gamma}   \frac{d \dot{x}^{\gamma}}{d s} \frac{d \dot{x}^{\mu}}{d s} \Bigg) \nn \\   &&  {\cal L}  g^{\mu \nu} \Bigg(2 g_{\mu \alpha, \gamma} \frac{d}{d s}  +  g_{\mu\alpha, \gamma, \delta}  \frac{d x^{\delta}}{d s} \Bigg)   \frac{d x^{\gamma}}{d s} \frac{d \dot{x}^{\mu}}{d s}  
\Bigg\}  \label{eq:geodsc1}
\eea
where the proper time can be deduced from the Lagrangian, $\tau = \int {\cal L}(s, \dot{x}, \ddot{x})\, ds$, 
\bea
\frac{\partial \tau}{\partial s} &=& {\cal L}(s, \dot{x}, \ddot{x}) = -\left[g_{\mu\nu} \left(\frac{d x^{\mu}}{d s} \frac{d x^{\nu}}{d s} + \beta \frac{d \dot{x}^{\mu}}{d s} \frac{d \dot{x}^{\nu}}{d s} \right)\right]^{1/2}, \nn \\
\Gamma^{\eta}_{\mu\nu} &=& \frac{1}{2} g^{\eta\alpha} \left[g_{\mu \alpha, \nu} - g_{\alpha \nu, \mu} + g_{\mu \nu, \alpha}\right], \nn \\
g_{\mu\nu, \alpha} &=& \frac{\partial g_{\mu \nu}}{\partial x^{\alpha}}, \nn \\
g_{\mu\alpha, \gamma, \delta} &=& \frac{\partial}{\partial x^{\delta}} \left(\frac{\partial g_{\mu \alpha}}{\partial x^{\gamma}}\right). \nn
\eea
\end{itemize}
The $\beta$--terms in Eq. (\ref{eq:geodsc1}) distinguish this expression of discretized geodesics from the GR geodesics, 
\bea
\frac{d^2 x^{\mu}}{d s^2} +\Gamma^{\eta}_{\mu\nu} \frac{d x^{\mu}}{d s}\, \frac{d x^{\nu}}{d s}=0,
\eea
which could be solved when, for instance, assuring that the proper time reads $ds^2=g_{\mu\nu} dx^{\mu} dx^{\nu}$. Under the condition that 
\bea
g_{\mu\nu} \frac{d x^{\mu}}{d s}\, \frac{d x^{\nu}}{d s} = \epsilon,
\eea
where $\epsilon=0$ for photon and $\epsilon=1$ for massive particles, the geodesic equation is supplemented. For conserved energy and angular momentum the geodesic equation can be expressed as an ordinary first--order differential equation \cite{Hackmann:2008zza}.

Solving the discretized geodesic equation, Eq. (\ref{eq:geodsc1}), is not a trivial task, because of the finite terms in rhs, which encompass higher--order derivatives.

\section{Conclusions}
\label{sec:cncls}

The approach of minimal length uncertainty conserves both Christoffel connection\footnote{The metric connection characterizes the affine connection to a manifold provided with the metric. This allows lengths, for instance, to be measured on that manifold.} and the equivalence principle of the gravitational and inertial mass. They are not affected. This allows to recover any violations in the equivalence principle, especially as a possible reason of quantum gravity. We highlight that this conclusion contradicts previous studies \cite{Ghosh:2013qra, Scardigli:2007bw}. The reason could be the rigorous procedure implemented in the present study. We have first determined the modified metric tensor. Accordingly, we have obtained the modified line element. Both are key principles of the proposed theory of discretized geometry. We could also derive the corresponding geodesics, in which the appearance of higher--order derivatives substantially distinguishes this expression from the GR geodesics. 

The modification to the line element is summarized in the term $\beta^2 \ddot{x}^\lambda \ddot{x}_\lambda$, which highlights the essential contributions added by the acceleration of the test particle. We conclude that the length discretization ensures that the line element wouldn't only expand but apparently accelerates. The metric tensor is also modified by the term $\beta \ddot{x}^\lambda \ddot{x}_\lambda$. The properties of the modified metric tensor has been elaborated.

We find that the modified geodesics is not only providing the acceleration of a test particle inside the gravitational field. Higher--order derivatives are also present \cite{Eager_2016}, namely the jerk, $\dddot{x}^{\mu}$, which in turn is derived from the time derivative of the acceleration. In contract to GR geodesics, solving the modified geodesics isn't a trivial task. The corresponding differential equations are complex and non--linear.

\section*{Acknowledgments}

AT is very grateful to organizers of sixteenth Marcel Grossmann meeting - MG16 for their kind invitation and financial support!


\bibliographystyle{ws-procs961x669}
\bibliography{ATawfikRefsMG16}

\end{document}